\documentclass[aps,floats,showpacs,amstex,amssymb,bm,tightenlines,notitlepage,hidelinks,floatfix,superscriptaddress,nofootinbib]{revtex4-2}

\usepackage{tcolorbox}
\usepackage[utf8]{inputenc}
\usepackage[cm]{fullpage}
\usepackage{amsmath}
\usepackage{amsthm}
\usepackage{amsfonts}
\usepackage{amscd}
\usepackage{bm}
\usepackage{epsfig}
\usepackage{amssymb}
\usepackage{tabularx}
\usepackage{longtable}
\usepackage{calligra}
\usepackage{enumerate}
\usepackage{hyperref}
\usepackage{tikzsymbols}
\usepackage{tabularx}
\usepackage{multirow}
\usepackage{mathrsfs}
\usepackage{cancel}
\usepackage[normalem]{ulem}
\def\j2{\mathbf{J}^2}
\def\a{\alpha}
\def\b{\beta}

\def\e{\epsilon}

\def\w{\omega}

\def\p{\partial}

\def\W{\mathcal{W}}

\begin{document}
\title{Parallel waves  in Einstein-non linear sigma models}

\author{B\'eatrice Bonga}
\email{bbonga@science.ru.nl}
\affiliation{Institute for Mathematics, Astrophysics and Particle Physics, Radboud University, 6525 AJ Nijmegen, The Netherlands}
\author{Gustavo Dotti}
\email{gdotti@famaf.unc.edu.ar}
\affiliation{Facultad de Matemática, Astronomía, F\'{\i}sica y Computaci\'on (FaMAF), Universidad Nacional de C\' ordoba  
and Instituto de F\'{\i}sica  Enrique Gaviola, CONICET. Ciudad Universitaria, (5000) C\'ordoba, Argentina.}

\begin{abstract}
	We study a family of solutions of Einstein-non linear sigma models with 
	$S^2$ and $SU(2) \sim S^3$ target manifolds. In the $S^2$ case, the solutions are smooth everywhere, free of conical singularities, and approach asymptotically the metric of a cosmic string, with a mass per length that is proportional to the absolute value of the winding number from topological spheres onto the target $S^2$. This gives an interesting example of a relation between a mass and a topological charge. The case with target $SU(2)$ generalizes the
	 stationary solution found in Eur. Phys. J. C (2021) 81:55 to parallel waves with a non-planar wavefront $\mathcal{W}$. 
	 We prove that these $\mathcal{W}$-fronted parallel waves are sub-quadratic in the classification in Class. Quant. Grav. \textbf{20} (2003) 2275, and thus causally well behaved. These spacetimes have a non-vanishing baryon current and their geometry has many striking features.
\end{abstract}
\maketitle

\tableofcontents

\section{Introduction: Einstein non-linear sigma model}
\label{sec:intro}

Quantum chromodynamics (QCD), a non-Abelian gauge theory with
$SU(3)$ gauge group, gives a description of hadrons in terms of their fundamental degrees of freedom:
quarks and gluons. Hadrons, being composite particles,   appear in the low energy limit of QCD, 
which corresponds to the non-perturbative, strongly coupled regime. At 
these energy scales it is found that effective theories become the most efficient tools to 
describe them. The leading term of the effective Lagrangian \textcolor{black}{in Minkowski spacetime} is  
the non-linear sigma model (NLSM) 
\begin{equation}\label{cl}
	\mathcal{L} = \frac{K}{4} \text{Tr} (L^a L_a),
\end{equation}
where we neglected the quark masses (see, e.g.,  Section VII.1 in \cite{DSM}) and 
\begin{equation}\label{MC}
	L_a = U^{-1}\p_a U 
\end{equation}
is the Maurer-Cartan form for a field  $U$  with target the group $SU(2)$. 
This effective theory  encodes the low energy dynamics of pions. \\

The NLSM \textcolor{black}{in Eq.~\eqref{MC}} cannot support static
solitonic solutions  in Minkowski spacetime. This 
was  proved long ago by Derrick using an elegant scaling argument \cite{8}. Solitons are of interest because they can be understood as baryons. Skyrme 
\cite{skyr} introduced a modification to the NLSM to allow for such static, topologically stable 
solitonic solutions  on Minkowski spacetime,  by adding to the Lagrangian in Eq.~\eqref{cl} the term $\text{Tr}([L_a,L_b][L^a,L^b])$, 
which  is \textit{part} of the sub-leading contributions to the 
QCD effective  Lagrangian (see section~XI.4 in \cite{DSM}, as well as chapter~9 in \cite{1}). \\

Derrick's scaling argument, however, can also be circumvented in other ways, since it uses the symmetries of
Minkowski spacetime and implicit boundary conditions. One such method is imposing periodic, crystal-like boundary conditions on flat spacetime
\cite{C1, C2}. Another, \textcolor{black}{and the one explored in this paper}, 
 is coupling the NLSM  to Einstein's gravity \cite{BT1, BT2, BT3, Canfora:2020ppn}.
In fact, if we \textcolor{black}{minimally} 
couple the NLSM to gravity, it is possible to find solitonic solutions keeping  only the low energy leading term  Eq.~\eqref{cl}, 
that is, working with the action 
\begin{equation}\label{action}
	S= \int_M d^4 x \sqrt{-g} \; \left(\frac{\mathcal{R}}{2 \kappa} +\frac{K}{4} \text{Tr} (L^a L_a) \right), 
\end{equation}
where $L_a$ is the Maurer-Cartan form in Eq.~(\ref{MC})  for a field $U: M \to SU(2)$, $(M,g_{ab})$ the spacetime, $\kappa$ is Newton's constant
and $K$ is the coupling constant of the
NLSM, which is proportional to the square root of  the decay constant of pions. In geometrized units, we have \cite{Canfora:2020ppn}
\begin{equation}\label{pc}
	0 < K \kappa \ll 1.
\end{equation}
The NLSM term in  (\ref{action})
 has an interesting geometric interpretation: \textcolor{black}{as we prove in Section \ref{Spws},} it is (proportional to) the trace of the pullback of the $S^3 =SU(2)$ metric onto spacetime. \textcolor{black}{
As such, it belongs to the family of Einstein-NLSM (ENLSM). In these theories,  
 there is field $\Psi: M \to N$ with target a compact boundaryless Riemannian 
manifold $(N,G_{AB})$, and the action is given by the Einstein-Hilbert term plus the trace of the pullback by $\Psi$ of the metric $G_{AB}$:
\begin{equation}\label{nlsm}
S_{ENLSM}= \int d^4 x \sqrt{-g} \; \left( \frac{\mathcal{R}}{2 \kappa} + 
 \frac{K}{2} g^{ab} \p_a \Psi^A \p_b \Psi^B G_{AB}(\Psi(x)) \right).
 \end{equation}
 }
Previous related work  \cite{BT1,BT2,BT3} uses the full Skyrmion model coupled to Einstein gravity instead of the simpler action in Eq.~\eqref{action} (or equivalently Eq.~\eqref{nlsm}).  
This simpler action was, however, recently considered in \cite{Canfora:2020ppn}, where solutions with a  stationary spacetime metric were found that describe 
solitonic matter. \\

In this work, we generalize those stationary spacetimes 
 to be dynamical. Specifically, we find solutions to the action in Eq.~(\ref{action}) for which the spacetime metric has a Kerr-Schild character and describes a  
 parallel wave with a non-planar wavefront with  transverse  metric $ds^2_\W$:
\begin{equation} \label{wfw}
	ds^2 = -du dv + H(u,\rho,\phi) du^2 + \underbrace{d\rho^2 + S(\rho)^2 d\phi^2}_{=ds^2_{\W}} .
\end{equation}
Metrics of this form generalize  plane-fronted waves with parallel propagation,  or pp-waves for short. Such generalized pp-waves were studied in \cite{Candela:2002rr,Flores:2002fx, Flores:2004dr}, where it was found that the rate of growth of  $H$ as a function of the distance $d$ to a fixed point on  $\W$, as $d \to \infty$, determines the causal behavior of the spacetime. 
We show below that our  solutions  correspond to the sub-quadratic case in the classification in \cite{Candela:2002rr,Flores:2002fx,Flores:2004dr}, which has a much better causal behavior than the ordinary, plane fronted pp-waves \cite{Penrose:remarkable}. These solutions are interesting not only given the physical model they are derived from, but also because of their geometric properties \textcolor{black}{as parallel waves traveling on a cylindrical background.\\}

\textcolor{black}{We also study a particular  static solution of the theory (\ref{nlsm}) with target $S^2$. This solution has a metric with 
 cylindrical symmetry, that is, 
of the form in Eq. (\ref{wfw}) with $H=0$. It is 
 an interesting  example of an \emph{everywhere smooth} metric, which \textit{asymptotically} looks like that of a cosmic string, since  
 $S(\rho) \simeq \b_+ 
 \rho $ for large $\rho$ with $0<\beta_+<1$. On the other hand,   near $\rho=0$ we find that  $S \simeq \rho + \mathcal{O} (\rho^2)$,
  which assures that $ds^2_\W$ is free 
 of conical singularities. Most interesting, 
 the mass per length $2\pi(1-\b_+)/\kappa$ of the asymptotically apparent string is  proportional to the
 absolute value of the winding number
 of -topologically- spacetime spheres onto the target $S^2$. \\}

The paper is organized as follows. \textcolor{black}{In Sec.~\ref{Spws} we  review the derivation of the baryon charge conservation 
of the theory (\ref{nlsm})} and discuss some subtleties about 
integration  on hypersurfaces that are not everywhere spacelike.  
In Sec.~\ref{sec:parallel-wave-solutions}, we present the \textcolor{black}{field equations 
of the action Eq. (\ref{nlsm})  and derive the solution with the metric as in Eq.~\eqref{wfw}. 
A by-product of these calculations gives a static solution of the theory (\ref{nlsm}) with target $S^2$. 
This is discussed 
 in Sec.~\ref{Sstbc}}, where 
 we elaborate on its geometry and establish the relationship between its mass per length and the topological winding number. 
 \textcolor{black}{ In Sec.~\ref{sec:non-vanishing-baryon-current} we return to the $SU(2)$ NLSM.} 
  After exploring its geometry through the study of geodesics,
 we discuss the baryon charge and, for the particular case in \cite{Canfora:2020ppn} of a non-static $U$ field leading to a static metric, 
we give  different notions of mass per length and \textcolor{black}{relate it 
to the baryon charge. The $SU(2)$ field configuration that we analyze does not carry a topological charge, 
and can be regarded as a parallel wave propagating in an otherwise cylindrical spacetime.} 
 We end with Sec.~\ref{sec:conclusion} by summarizing our main results and discussing their implications.

\section{Fields and conserved currents}\label{Spws}
In this section, we derive the equation for the baryon conserved current thereby providing some further physical background to the 
\textcolor{black}{QCD Einstein-NLSM 
with action (\ref{action}).}
We also explain in some detail how to calculate the total baryon charge at an open  spacelike surface $\Sigma'$ that is 
computationally challenging to find, by using Stokes/Gauss  theorem and replacing
it with an integral on an asymptotically matching surface $\Sigma$. \\

We parametrize
$SU(2) \sim S^3$ using  hyper-spherical coordinates $z^A=(\a, \Theta$, $\Phi)$:
\begin{equation}\label{s3}
\mathbb{R}^4 \ni \left( \begin{array}{c} x^1 \\ x^2 \\x^3 \\ x^4 \end{array} \right) = 
\left( \begin{array}{c} \sin \a \sin \Theta \sin \Phi \\ \sin \a \sin \Theta \cos \Phi \\ \sin \a \cos \Theta \\ \cos \a \end{array} \right).
\end{equation}
The normalized $S^3$ metric $G_{AB}$ is 
\begin{equation}\label{s3m}
G_{AB} dz^A dz^B = \frac{1}{2\pi^2} \left( d \a^2 + \sin^2 \a  d\Theta^2 + \sin^2 \a  \sin^2 \Theta d \Phi^2 \right).
\end{equation}
In terms of the coordinates $(\a,\Theta,\Phi)$, the $SU(2)$ matrices are given by 
\begin{equation}\label{U}
U^{\pm 1}= \cos(\a) \;\mathbf{1}_2 \pm \sin(\a) \; \hat n \cdot \mathbf{ t} = e^{\pm \a \hat n \cdot \mathbf{t}}, \;\;\, \mathbf{t}=(i \sigma_1, i \sigma_2, i \sigma_3)
\end{equation}
with $\sigma_j$ the Pauli matrices, 
\begin{equation}\label{pauli}
\sigma_1 = \left( \begin{array}{cc} 0&1\\1&0\end{array}\right), \;\;\; 
\sigma_2 = \left( \begin{array}{cc} 0&-i\\i&0\end{array}\right), \;\;\; 
\sigma_3 = \left( \begin{array}{cc} 1&0\\0&-1\end{array}\right)
\end{equation}
and 
\begin{equation}\label{n}
\hat n =( \sin(\Theta) \cos(\Phi), \sin(\Theta) \sin(\Phi),  \cos(\Theta)).
\end{equation}
The three pion fields are --mod normalization conventions--  $\bm{\pi} = \alpha \hat{n}$ \cite{DSM}.
Inserting Eqs.~(\ref{U})-(\ref{n}) in Eq.~(\ref{MC})  gives 
\begin{equation}
L_a = \left[ (\p_a \a) \; \hat n + \sin( \a )\cos( \a) \;  \p_a \hat n + \sin^2(\a)  \left( \hat n \times \p_a \hat n \right) \right] \cdot \mathbf{t}
\end{equation}
This implies,  \textcolor{black}{as anticipated}, that 
\begin{equation}\label{pullback}
-\tfrac{1}{2} \text{Tr} [L_a L_b] =   \p_a \a \, \p_b \a + \sin^2 \a \;\p_a  \Theta \, \p_b \Theta + \sin^2 \a \sin^2 \Theta \; \p_a \Phi\,  \p_b \Phi
\end{equation}
is the pullback onto spacetime of the $S^3$ metric in Eq.~(\ref{s3m}) \textcolor{black}{ (compare (\ref{pullback}) with (\ref{s3m}))}, whose trace 
 $\text{Tr} [L^a L_a]$ appears in the action (see Eq.~\eqref{action}). \\
 
 The vector field 
 \begin{equation}\label{bc}
 J^a = \tfrac{1}{24 \pi^2} \e^{b c d a} \text{Tr} (L_b L_c L_d)
 \end{equation}
 describes the baryon current and is dual to the 3-form (conventions as in \cite{wald})
  \begin{equation}\label{bc2}
 {}^*J_{bcd} = J^a \e_{abcd}=
  \tfrac{1}{4 \pi^2}  \text{Tr} (L_{[b }L_c L_{d]}) = \tfrac{3}{ \pi^2} \sin^2 (\a) \sin (\Theta) \; \p_{[b }\a \;  \p_c \Theta \; \p_{d]} \Phi .
 \end{equation}
 Index anti-symmetrization is defined as sum over signed permutations 
 divided by the factorial of the number of anti-symmetrized indices, so the above equation 
 can be written in the language of forms as 
  \begin{equation}\label{bc3}
  \tfrac{1}{24 \pi^2}  \text{Tr} (L \wedge L \wedge L) = \tfrac{1}{2\pi^2} \sin^2 \a \sin^2 \Theta \; d\a \wedge d \Theta \wedge d\Phi,
 \end{equation}
which is the pullback of the normalized $S^3$ volume form from Eq.~(\ref{s3m}). Since exterior derivatives commute with 
pullbacks and the dimension of $S^3$ is 3,  the exterior derivative of this 3-form vanishes. 
 In view of the duality in Eq.~(\ref{bc2}), this is equivalent to the condition of \textit{baryon current conservation}:
 \begin{equation}\label{bcc}
 \nabla_a J^a =0.
 \end{equation}
The total  baryon charge $B_{\Sigma'}$ measured by the observers with velocities $n^a$ 
 normal to an (open oriented) spacelike hypersurface $\Sigma'$ is 
\begin{equation}\label{se0}
B_{\Sigma'}=   \int_{\Sigma'} J_a n^a\; \e^{\Sigma'}_{bcd}, \;\;\; (\e^{\Sigma'}_{bcd} = \e_{abcd} n^a). 
  \end{equation}
If $B_{\Sigma''}$ is a second such a surface, and either $\Sigma'' - \Sigma' = \p V$, the boundary of  an open subset of spacetime 
(the relative sign here indicates reversal 
of the normal), 
or $V$ is topologically a cylinder with cups $\Sigma''$ and  $\Sigma'$ and the fields decay fast enough so that the flow 
through the lateral is null, then 
charge is conserved, meaning that $B_{\Sigma'}=B_{\Sigma''}$. This follows from Gauss theorem and Eq.~(\ref{bcc}), or equivalently to the 
dual Stokes theorem and the fact that the 3-form dual to $J_a$, Eq. (\ref{bc2}),
 is closed. It is important to recall how one switches from Stokes' to Gauss' version 
in the most general case: 
 Consider a --possibly non closed-- orientable hypersurface $\Sigma \subset M$, choose a normal smooth vector field $N^a$ on $\Sigma$. 
 We allow the case where  $\Sigma$  changes character from 
 spacelike to timelike, as long as its normal $N^a$ is null only on a zero measure set $\Sigma_o$. Let $n^a=N^a/\sqrt{|N^cN_c|}$. This 
 vector field is smooth on 
 $\tilde \Sigma =\Sigma \setminus \Sigma_o$ (a disconnected set if $\Sigma_o \neq \emptyset$), 
 undefined  on $\Sigma_o$, and it is normalized to $n^a n_a =-1$ ($+1$) on the spacelike (timelike) sectors 
 of $\Sigma$. Let $\e^{\tilde \Sigma}_{bcd} = \e_{abcd}n^a$ be the volume form on $\tilde \Sigma$. Given a 3-form $\a_{abc}$ on $M$ with  dual 
 $v_a =\tfrac{1}{6}\e_{b c d a} \a^{b c d}$  (that is, $\a_{abc}= \e_{dabc} v^d$), we have
 \begin{equation}\label{se}
 \a_{bcd}\mid_{\tilde \Sigma}  = -v_a n^a (n^kn_k) \e^{\tilde \Sigma}_{bcd} \mid_{\tilde \Sigma},
 \end{equation}
 where on the left side we mean pullback. 
 The above equality is used when proving Gauss theorem from Stokes theorem on manifolds with boundary (the $n^kn_k=\pm 1$ factor 
 is the reason why we need to switch  from outer to inner normal  when leaving  spacelike  sectors of the boundary). 
  Equation (\ref{se}) is 
 particularly useful when it is difficult to explicitly determine the timelike/spacelike sectors of $\Sigma$, since its left side is insensitive 
 to these changes. 
 From now on we will treat  \textit{integrals} on $\Sigma$,  thus the distinction between  $\Sigma$ and $\tilde \Sigma$ is irrelevant. 
 Suppose $\Sigma$ is an open  hypersurface that is asymptotically spacelike, and that can be deformed onto a  hypersurface 
 $\Sigma'$ that is spacelike everywhere and 
 matches $\Sigma$ in the asymptotic region.  Suppose we are interested in the total  charge 
  $\int_{\Sigma'} v_a n^a \e_{\Sigma'}$ for a divergence-free vector field $v^a$, 
 we do not know  $\Sigma'$ in detail and, although we do know $\Sigma$, we would like to avoid  determining the sectors where 
  $\Sigma$ is timelike/spacelike. In this case we can use Gauss theorem and Eq.~(\ref{se}) and find that (the orientation is chosen such that, when 
  timelike, $n^a$ is  future pointing)
  \begin{equation}\label{se3}
  \int_{\Sigma'} v_a n^a\; \e^{\Sigma'}_{bcd} = \int_{\Sigma} -v_a n^a  (n^k n_k) \; \e^{\Sigma}_{bcd} = \int_\Sigma  \e_{abcd} v^a .
  \end{equation}
  For $v^a=J^a$ given in Eq.~(\ref{bc}) the above equation gives
   \begin{equation}\label{se2}
B_{\Sigma'}=   \int_{\Sigma'} J_a n^a\; \e^{\Sigma'}_{bcd} = \frac{1}{4\pi^2}\int_\Sigma \text{Tr}(L_{[a}L_bL_{c]}) .
  \end{equation}
  The integral on the left is the total baryon charge measured by observers with velocity $n^a$. 
  The above equation shows that this can be calculated  as an integral over the asymptotically matching surface 
  $\Sigma$ without  knowing the sectors where  $\Sigma$ 
  is not spacelike. \\
  
  The baryon charge in Eq.~(\ref{se2}) is a conserved quantity in the sense that the integral on the right  is the same on 
  surfaces in the same homology class. This quantity, however, does not necessarily have a \textit{topological meaning}. 
  If  $\Sigma$ is open and complete \textit{and the field $U$ tends to a constant},  
  (say, the identity matrix) in the asymptotic region, we may regard $B_{\Sigma'}$ as the integral of the pullback of the normalized $S^3$ metric onto a manifold that is topologically equivalent to $S^3$ (the one point compactification 
  of $\Sigma'$). In this case $B_{\Sigma'}$ will be an integer: the number of times this sphere wraps around $SU(2) = S^3$. In general, however, 
  $U$ does not have a common limit at infinity, and thus 
  $B_{\Sigma'}$ is \textit{not}  an integer.

\section{Parallel wave solutions}\label{sec:parallel-wave-solutions}
In this section, we present the field equations \textcolor{black}{of the QCD ENLSM (\ref{action})} and a general class of parallel 
wave solutions that generalize the solutions with static metrics found  in \cite{Canfora:2020ppn}.
The field equations derived from the action in Eq.~(\ref{action}) are
\begin{equation}\label{su2eq}
\nabla^a L_a =0,
\end{equation}
and 
\begin{equation} \label{ee}
G_{ab} = \kappa T_{ab},
\end{equation}
where the  energy-momentum tensor of the $SU(2)$ field is 
\begin{equation}\label{tab3}
\begin{split}
T_{a b } = -\frac{K}{2} \text{Tr} (L_a L_b  - \tfrac{1}{2}g_{a b} L^c L_c) &=
 K \left[\sin^2(\a) \sin^2(\Theta)  \left(\p_a \Phi \p_b \Phi   -\tfrac{1}{2} g_{a b} \p^c \Phi \p_c \Phi \right)  \right. \\ 
 & \quad \left.  + \sin^2(\a) ( \p_a \Theta \p_b \Theta -\tfrac{1}{2} g_{a b} \p^c \Theta \p_c \Theta)+  
(\p_a \a \p_b \a -\tfrac{1}{2} g_{a b} \p^\gamma \a \p_\gamma \a) \right].
\end{split}
\end{equation}
Eq.~(\ref{su2eq}) shows a minimal coupling to gravity of the pion field equation in Minkowski spacetime. 
Note that, since $L_a=U^{-1} \p_a U$, this 
equation is indeed second order in the pion fields $\bm{\pi}$.  Note also that it has the form of a conserved current equation. This is  because the action in 
Eq.~(\ref{action}) is invariant 
under the $SU(2) \times SU(2)$ (global) transformation
\begin{equation}\label{LR} 
U \to g_L U g_R^\dagger, \;\;\; (g_L,g_R) \in SU(2) \times SU(2),
  \end{equation}
and $L_a$ is the Noether current under the left $SU(2)$ subgroup for which $g_R=\mathbf{I}$ \cite{Scherer:2002tk}. The conservation of the additional conserved current from the subgroup $g_L=\mathbf{I}$ is trivially related to Eq.~(\ref{su2eq}). \\

The field equations  (\ref{su2eq})-(\ref{ee}) 
admit solutions in which the metric is a     $\W$-fronted parallel wave metric (as defined in  \cite{LNP} and references therein):
\begin{equation}\label{background}
ds^2 =-dt^2+dz^2 + H(t-z,r,\phi) (dt-dz)^2 +\underbrace{\ell^2  \; e^{-2R(r)} \; (dr^2 + d \phi^2)}_{=ds^2_{\mathcal{W}}}. \\
\end{equation}
We may occasionally switch to null coordinates in $t-z$ space:
\begin{equation} \label{variables}
u = t-z, \;\;\; v =t+z
\end{equation}
and use an alternative radial variable $\rho$ for the wavefront cross section $\mathcal{W}$, which has 
$(+,+)$ metric 
\begin{equation}\label{arv}
ds^2_{\mathcal{W}}=\ell^2  \; e^{-2R(r)} \; (dr^2 + d \phi^2) = d \rho^2 +S^2(\rho) d\phi^2.
\end{equation}
In terms of $r$,  $\rho$ and $S(\rho)$ are given by
\begin{equation}\label{rho0}
\frac{d\rho}{dr} = \pm S, \qquad   S = \ell e^{-R(r)} .
\end{equation}
Switching to  $(u,v,\rho,\phi)$ coordinates, the metric in Eq.~(\ref{background}) becomes that of Eq.~(\ref{wfw}). \\

Note that $H, R, r$ and $\phi$ are dimensionless, $\ell, t, z, u, v, \rho$ have dimensions of length, and 
\begin{equation}
 \; \; -\infty < r,z,t, u,v < \infty, \;\; \phi \sim \phi+2 \pi .
\end{equation}
The range of $\rho$ depends on how $S$ decays as $r \to \pm \infty$. \\

The wave vector vector $k^a$, given in  $(u,v,*,*)$ coordinates by $k^a \p_a = \p_v$ is: i)  orthogonal to the wavefronts $u=$ constant, ii) null
and iii) covariantly constant. The latter property ensures that the spacetime is a member of the Kundt class, which are Lorentzian manifolds 
admitting a geodesic null 
congruence with vanishing optical scalars (expansion, shear and twist) \footnote{That the metric is Lorentzian irrespective 
of the range of $H$,  is clear from the fact that the determinant of (\ref{background})  
 is  $- \ell^4 \exp(-4R(r)) <0$.}.  The wave vector $k^a$ is used to select a time orientation by defining it to be future oriented. This choice implies that,  
in those regions where $\p_t$  in Eq.~(\ref{background}) is timelike,  it is future oriented. We call these 
spacetimes $\W-$fronted parallel waves because the wave vector $k$ is covariantly constant and the transverse 
metric on the wavefronts is $ds^2_\W$. 
These generalize pp-waves, which correspond to the particular case where $ds^2_\W$ is planar. 
We will prove below that the decay of $H$ at large distances along $\W$ guarantees that our solutions fall, in the classification in  
\cite{Flores:2002fx}, in the subquadratic type, making them 
  causally well behaved, contrary to what happens  with the flat fronted pp-waves \cite{Penrose:remarkable}. \\

Our $SU(2)$ field ansatz is
\begin{equation}\label{su2field}
\a=\a(r), \;\;\; \Theta= q \phi, \;\;\; \Phi=F(t-z),
\end{equation}
where $F$ is a function that models the wave profile. 
It generalizes  that in \cite{Canfora:2020ppn}  allowing for non-stationary spacetime metrics 
\footnote{The solutions of the system in Eq.~(\ref{action}) 
presented in \cite{Canfora:2020ppn} correspond to $F=\w(t-z)$ and lead to a static metric with $H(t-z,r,\phi)=H(r,\phi)$ in (\ref{background}).}.

For the ansatz (\ref{background})-(\ref{su2field}), we find that Eqs.~(\ref{su2eq})-(\ref{ee}) reduce to three independent field equations: 
\begin{equation}
\begin{split} \label{aR}
\big( \a'(r)\big)^2 &= q^2 \; \sin^2(\a(r)),\\ 
R''(r) &= K \kappa q^2\sin^2(\a(r)),
\end{split}
\end{equation}
and (note the trivial way $F'(t-z)$ appears in this equation):
\begin{equation}\label{Heq}
\begin{split}
(\p_r^2 +\p_\phi ^2) H &= - 2 K \kappa \ell^2 \exp(-2R(r)) \sin^2(\a(r)) \,\sin^2(q \phi) (F'(t-z))^2
 \\& = - 2  \left(\frac{\ell }{q}\right)^2 \exp(-2R(r)) R''(r) \,\sin^2(q \phi) (F'(t-z))^2.
\end{split}
\end{equation}
We find that the function $F$ is neither constrained by  the field equations nor by energy conditions. The 
dominant and strong energy conditions are satisfied in any case for this theory, as proved in  \cite{Gibbons:2003cp}. 
Alternatively, the strong energy condition
\begin{equation}
	R_{ab} \zeta^a \zeta^b \geq 0 \;\; \text{ for  timelike  } \zeta^a.
\end{equation}
follows from  Proposition 2.2 in \cite{Candela:2002rr} and the facts 
  that Eq.~(\ref{Heq}) implies that the $\mathcal{W}$-Laplacian of $H$ is negative, whereas Eq.~(\ref{aR}), together with
  $R^{\mathcal{W}}_{jk}=\text{diag} (d^2R/dr^2, d^2R/dr^2)$ in $(r,\phi)$ coordinates,  
imply that $R^{\mathcal{W}}_{jk}$ is positive definite. \\

Discarding the uninteresting  solution to Eq.~\eqref{aR} with $\alpha(r)=0,\pi$ and $R(r)=Ar+b$,  we are left with the solution
\begin{equation}\label{gs}
\begin{split}
\alpha(r) &= 2 \arctan \left( \exp (\e |q|r+C_1) \right),\\
R(r)& = K \kappa \ln \left( \cosh(\e |q|r+C_1) \right) +C_2r+C_3 ,
\end{split}
\end{equation}
with $\e=\pm 1$ and $C_1,C_2$ and $C_3$ arbitrary constants.
Interestingly, despite the fact that the field equations are non-linear and coupled in an intricate  way,  
\textit{whether or not} $F'$ is a constant (and consequently whether the metric is static or dynamical), 
\textit{the field equations for the non-linear sigma model together with the corresponding Einstein
equations lead to the same solutions for}  $\alpha (r)$ \textit{and} $R(r)$. 
This is a remarkable property of the ansatz in Eqs.~\eqref{background} and \eqref{su2field} that allows us to disentangle the 
physical effects introduced through $F(u)$.\\

We now analyze the constraints on  the relevant integration constants. 
All  algebraic  (that is, non-differential) curvature scalar fields made  out of the Riemann tensor, the metric and its inverse, and the volume form
-- for which a basis is given in 
\cite{invars} --  are powers of the Ricci scalar $\mathcal{R}$:
\begin{equation}\label{cs}
 \mathcal{R} = \frac{2}{\ell^2}\; e^{2R(r)} R''(r)
\end{equation}
and thus are independent of $H$. Since 
\begin{equation}
e^{2 R(r)} R''(r) = 
K \kappa q^2 \cosh(\e |q| r+C_1)^{2K\kappa-2} e^{2C_2r+2C_3}  
\sim e^{|q||r|(2K\kappa-2)} e^{2C_2r} \qquad \text{ as } \; |r| \to \infty,
\end{equation}
in view of (\ref{pc}), requiring that 
the scalars of curvature be well behaved for $- \infty < r <\infty$ is equivalent to the condition found in    \cite{Canfora:2020ppn}: 
\begin{equation}\label{cond}
|C_2|  < (1-K \kappa) |q|.
\end{equation}

The  relation between the radial coordinates $-\infty<r<\infty$ and $\rho$ is $d\rho/dr =\pm S$, with 
 \begin{equation} \label{l1}
S =
\ell \,  e^{-R(r)} = \ell  \cosh(\e |q| r+C_1)^{-K\kappa} e^{-C_2r-C_3}\simeq \nu_{\pm} e^{-K \kappa |q||r|} e^{-C_2r} 
  =:    \nu_\pm \, e^{\b_{\pm} r} \qquad \text{as}  \; \; 
 r \to \pm\infty, 
 \end{equation}
 where we assumed that the constants 
 \begin{equation}\label{betas}
 \b_+ = -K \kappa |q|-C_2, \;\; \beta_-= K\kappa |q| -C_2
 \end{equation}
 are non-zero. The positive constants $\nu_\pm$ depend on $C_1, C_3, \epsilon, \kappa K$ and $\ell$. 
 Note that 
 \begin{equation}
\b_- -\b_+= 2 K \kappa |q| \geq 0.
\end{equation}

This inequality allows three out of four sign possibilities: i) $\b_-$ and $\b_+$ are positive, ii) 
$\b_-$ is positive and $\b_+$ negative, iii) 
$\b_-$ and $ \b_+$ are negative.
 The solutions of type iii) are trivially related to those of type i). This is a consequence of the symmetry of the metric (see Eq.~(\ref{background})) under $r \to -r$:
 given a solution $\a(r)$, $R(r)$ and $H(t,z,r,\phi)$ of the field equations (\ref{aR})-(\ref{Heq}), 
the functions  $\tilde \a(r)=\a(-r)$, $\tilde R(r)=R(-r)$ and $\tilde H(t,z,r,\phi)=H(t,z,-r,\phi)$ is also a solution, 
and the asymptotic behaviors of these two solutions are related by $(\tilde \b_-, \tilde \b_+)= ( - \b_+,-\b_- )$. 
 We will then assume from now, without loss of generality,  that $\b_->0$. This guarantees  
that  $r=-\infty$ is \textit{a point} at a finite distance  from any other point in $\W$. We 
define $\rho$ to be the $\W-$geodesic distance to this point:
 \begin{equation} \label{rho}
 \rho(r) = \ell \int_{-\infty}^r e^{-R(y)} dy,  
 \end{equation}
 so that   $\rho \to 0$ as $r \to -\infty$,  the upper sign choice holds in Eq.~(\ref{rho0}), that is $d\rho/dr=S=\ell e^{-R(r)}$, and 
 \begin{equation}
 S(\rho) \simeq \b_- \rho \;\;\text{ as } \rho \to 0 . 
 \end{equation}

 To avoid conical singularities in $\W$, we further impose that $\b_-=1$. Adding also the regularity condition (\ref{cond}), the cases 
 of interest  narrow down to:
 \begin{itemize}
 \item Case 1: $\beta_-=1, \beta_+>0$. \\
 
 The values of the different constants are
 \begin{equation} \label{B}
\beta_-=1, \;\;\;  \beta_+ =  1-2\kappa K |q|, \;\;\; 1<|q| \leq\frac{1}{2\kappa K}, \;\;\;  C_2 =  \kappa K |q|-1.
 \end{equation}
  $\W$ has the manifold structure of a plane with $(\rho,\phi)$ polar coordinates. The metric $ds^2_\W$ is regular 
  everywhere and asymptotically conical, with a  deficit angle at infinity of $2\pi(1-\beta_+)=4\pi K \kappa |q|$: 
 \begin{equation} 
 ds_{\mathcal{W}}^2 
 \simeq \begin{cases} d \rho^2+ \rho^2 \; d\phi^2 & \qquad \text{as}  \; \;   \rho \to 0 \\  d \rho^2+  (1-2\kappa K |q|)^2 \rho^2 \; d\phi^2 & \qquad \text{as}   \; \; \rho \to \infty.
 \end{cases}
 \end{equation}
 The asymptotic formulas for the inverse of Eq.~(\ref{rho}) are:
\begin{equation}
r \simeq 
	\begin{cases} 
		\ln(\rho/\ell) &\qquad  \text{ as } \; \;  \rho \to 0  \\
		\frac{1}{\b_+} \ln(\rho/\ell)  &\qquad  \text{ as } \; \;  \rho \to \infty .		
	\end{cases}
\end{equation}
 \item Case 2:  $\beta_-=1, \beta_+=-1$\\
 
The values of the different constants are
 \begin{equation} \label{C}
\beta_- =1, \;\;\; \beta_+=-1, \;\;\; |q| =\frac{1}{\kappa K}, \;\;\;  C_2 =  0.
 \end{equation}
This case is of little interest, as it requires fine-tuning: $K \kappa |q|=1$.  
Let 
  \begin{equation} 
 \rho_\infty = \ell \int_{-\infty}^\infty e^{-R(y)} dy, 
 \end{equation}
then  $\W$ has the manifold structure of $S^2$ with  $\left(2\pi \frac{\rho}{\rho_\infty},\phi \right)$ 
  angular coordinates (respectively co-latitude and azimuth).  The sphere is equipped with a smooth metric,   
 smoothness at the poles follows from  
 \begin{equation} 
 ds_{\mathcal{W}}^2 
 \simeq 
 \begin{cases} 
 	d \rho^2+ \rho^2 \; d\phi^2 & \qquad \text{ as } \; \;   \rho \to 0 \\ 
 	d \tilde \rho^2+  \tilde \rho^2 \; d\phi^2 &\qquad \text{ as } \; \;   \ \tilde \rho\equiv \rho_\infty-\rho \to 0.
 \end{cases}
 \end{equation}
 The asymptotic formulas for the inverse of Eq.~(\ref{rho})  are
 \begin{equation}
r \simeq 
	\begin{cases} 
		\ln(\rho/\ell) & \qquad \text{ as } \; \;   \rho \to 0 \\
		- \ln \left( \frac{\rho_\infty-\rho}{\ell} \right) &  \qquad \text{ as } \; \; \rho  \to \rho_\infty.
	\end{cases}
\end{equation}

 \end{itemize}
 
The solution presented here, with waves traveling along the positive $z$-direction, could have been taken 
as traveling oppositely by proposing $F(v)$ instead of $F(u)$ in Eq.~\eqref{su2field}. A linear superposition of such waves does not lead to  solutions 
of the field equations. \\

\textcolor{black}{Note that we have solved two out of three field equations, those in (\ref{aR}). We postpone the treatment of the non-homogeneous linear 
equation (\ref{Heq}) to Sec.~\ref{sec:non-vanishing-baryon-current} and consider, in the following section, the trivial case where $F=H=0$.}

\section{An ENLSM with target $S^2$}\label{Sstbc}

The field equations (\ref{aR}) and (\ref{Heq}) admit the  solution $F=0$, $H=0$, with  $\a(r)$ and $R(r)$ as in Eq.~(\ref{gs}). 
This may at first look as an uninteresting solution, since $\Phi=0$ implies that  the baryon current vanishes (see Eqs.~(\ref{bc})-(\ref{bc2})).
\textcolor{black}{The $U$ field wraps around the $S^2$ equator of $S^3$}   
  defined by $(x^2)^2+(x^3)^2+(x^4)^2=1$ in  Eq.~(\ref{s3}):
\begin{equation}\label{s2}
\mathbb{R}^4 \ni \left( \begin{array}{c} x^1 \\ x^2 \\x^3 \\ x^4 \end{array} \right) = 
\left( \begin{array}{c} 0 \\ \sin \a \sin \Theta  \\ \sin \a \cos \Theta \\ \cos \a \end{array} \right).
\end{equation}
\textcolor{black}{This static solution of the QCD ENLSM (\ref{action}) is unstable since $U$ can unwrap 
within $S^3$ \footnote{We thank an anonymous referee for this observation.}. The instability can readily be checked: 
if we linearly perturb this solution  within
the $SU(3)$ theory by setting   $\a=\a(r)+ \e \a_1$, 
$\Theta=q \phi + \e \Theta_1$, $\Phi=\epsilon \Phi_1$ and keeping  only first order terms in $\e$, it readily follows 
from Eq. (31)-(32) that a possible solution is $\a_1 = 0$, $\Theta_1 = 0$, $H = 0$ and $\Phi_1$ an \textit{arbitrary} 
function of $t-z$ (the lack of backreaction is due to the fact that the right side of equation (32) is order $\e^2$). 
This certainly signals an instability, as the perturbation does not stay bounded in time, oscillating around the unperturbed static 
solution, as would be the case if this solution were stable.\\}

\textcolor{black}{However,  $\a(r)$ and $R(r)$ as in Eq. \eqref{gs} give a solution to a different theory: the ENLSM (\ref{nlsm}) with target $S^2$, 
the target 2-sphere being that defined in (\ref{s2}), parametrized with polar and azimuthal angles $\a$ and $\Theta$ respectively.}  This follows from the fact that
 for $\Phi \equiv 0$, the matter field piece in (\ref{action}) is the trace 
of the pullback of the $S^2$ metric, as follows from \eqref{pullback}, so in particular (\ref{gs}) is a stationary point of the action 
\begin{equation}\label{action2}
	\tilde S= \int_M d^4 x \sqrt{-g} \; \left(\frac{\mathcal{R}}{2 \kappa} -\frac{K}{2} g^{ab} \left( \p_a \a \p_b \a +\sin^2 \a \; \p_a \Theta \p_b \Theta
	 \right) \right)
\end{equation}
for the $S^2$ ENLSM. \\

The metric in this case has cylindrical symmetry:
\begin{equation}\label{background0}
ds^2=-dt^2+dz^2 +d\rho^2 + S^2(\rho) \, d\phi^2
\end{equation}
and belongs to the class of Petrov type~D spacetimes. 
The vector field  $t^a \p_a= \p_t$ is a timelike global Killing vector field, orthogonal to the 3-Riemannian slices with metric 
$dz^2 +d\rho^2 + S^2(\rho) \, d\phi^2$. 
The results in Section~\ref{sec:geom} show that this metric is geodesically complete. 
In Case 1, defined in Eq. \eqref{B}, we get an everywhere smooth solution, free of conical singularities, which   asymptotically looks 
like a cosmic string presenting a deficit angle sourced by regular matter fields. In Case 2, Eq. \eqref{C}, the $t=$ constant slices 
are cylinders $S^2_{(\rho,\phi)} \times \mathbb{R}_z$. \\

There is a topological number $q \in \pi_2(S^2) =\mathbb{Z}$ associated to these solutions, \textcolor{black}{which guarantees their stability 
as  solutions of the $S^2$ ENLSM.} Its absolute value is proportional to the mass per length, as we now proceed to prove.

\subsection{Topological number}

In view of the first equation in \eqref{gs}, $\a(r)$ covers monotonically the interval $(0,\pi)$ as $r$ goes from minus to plus infinity. This assures that 
(assuming $q$ is an integer) the map from the $t=t_o,z=z_o$ submanifolds $\W$  onto $S^2$ are well defined in Case 2 
(for which $\W$ is a sphere). Moreover,  in Case 1, for which $\W$ is a plane, the limits at infinity are direction independent, so we get 
 a map of the one point compactification of this plane, which is topologically a 2-sphere. As a consequence, in either case we have a topological 
number associated to this map. 
To compute it, we note that the 
 canonical $S^2$ metric $d \alpha^2+ \sin^2 \alpha \; d\Theta^2$ has normalized volume form $\tfrac{1}{4\pi} \sin\a \;  d\a \wedge d \Theta$ 
which pulls back to the spacetime 2-form  
\begin{equation} \label{qdensity}
\w_q=\tfrac{q}{4\pi }\sin (\alpha (\rho))(d\alpha /d\rho )d\rho \wedge d\phi . 
\end{equation}
 Since $\w_q$ is closed, its integral on any two-surface $\mathcal{W}'$ in the same homology class as a $t=t_o$, $z=z_o$ two-surface $\mathcal{W}$ gives 
\begin{equation}\label{wn}
\int_{\mathcal{W}'} \w_q =-(q/2)[\Delta \cos (\alpha )]= \e \, q. 
\end{equation}
This is  the -signed- number of times that $\mathcal{W}'$ wraps around the target 
$S^2$ (that is, $\e q \in \pi_2(S^2)$) . \\

\subsection{Mass per length}\label{mf0}

For the spacetime metric in Eq.~\eqref{background0}, we find  
\begin{equation}\label{Tab}
T_{ab} =\frac{1}{\kappa} G_{ab} = \frac{1}{2 \kappa} \; \text{diag}(\mathcal{R}_\W,-\mathcal{R}_\W,0,0) \equiv (e,-e,0,0)
\end{equation}
and 
\begin{equation}
R^a{}_b =  \frac{1}{2}\text{diag}(0,0,\mathcal{R}_\W,\mathcal{R}_\W) 
\end{equation}
where $\mathcal{R}_\W=-2S''(\rho)/S(\rho)$ is the Ricci scalar of 
$ds^2_\W=d\rho^2+ S^2(\rho) \, d\phi^2$.\\

 Like Minkowski spacetime, the metric (\ref{background0}) has a unit norm timelike covariantly constant vector field $t^a \p_a=\p_t$, 
 orthogonal to $t=$constant hypersurfaces, which can be regarded as a velocity field of the congruence of 
 privileged,  ``inertial'' observers. The current $J^a=-T^a{}_b t^b$ (4-momentum 
 density measured by these observers) is conserved: $\nabla_a J^a=0$. Its flow through 
 a $t=$constant surface $\Sigma$ gives the total energy measured by these observers, and this is  conserved in time.
 The volume form on $\Sigma$ is   $\epsilon_\Sigma = S(\rho) \; d\rho \wedge d\phi \wedge dz$,  the normal is $t^a$, so that  
 we need to  integrate   $e  \varepsilon_\Sigma = -\tfrac{1}{\kappa}S''(\rho) \; d\rho \wedge d\phi \wedge dz$ to obtain the total energy. 
 The mass per unit length on $\Sigma$ is obtained by omitting the integration on $z$, and is found to be proportional 
 to the absolute value of the topological charge (\ref{wn}):
 \begin{equation}\label{mu}
 \mu = -\frac{1}{\kappa} \int_{\mathcal{W}} S''(\rho) \; d\rho \wedge d\phi = \frac{2\pi}{\kappa} (1-\beta_+) = 4 \pi K |q|  .
 \end{equation}
 Note that $\mu  = \frac{2\pi}{\kappa} (1-\beta_+) $ is a standard result for cosmic strings \cite{linet}. \\

We conclude that  this simple solution  of the $S^2$ ENLSM theory is: 
i) smooth everywhere, 
ii) geodesically complete, 
iii)  free of conical singularities, 
iv) asymptotically conical in Case 1, 
with a mass per length sourced on the NLSM and proportional to the (absolute value) of its topological charge.\\

\textit{Remark.} For electromagnetic fields, there is a direct link between the vanishing of the magnetic part of the Weyl tensor and the vanishing of the vorticity 
 tensor $\omega_{ab}$ of the time translation Killing vector field (i.e., $\omega_{ab} = - \nabla_{[a} t_{b]}+ a_{[a} t
 _{b]}$ with the acceleration given by $a_a = t^b \nabla_b t_a$) \cite{Herrera:2006cw}.  There are no such general results known for the ENLSM, 
 but this example illustrates that this link in the electromagnetic case might be more general, as we find that 
the electric and magnetic part of the Weyl tensor in $(t,z,\rho,\phi)$ coordinates are
 \begin{align}
 	\mathcal{E}_{ab} &:= C_{acbd} t^c t^d = \text{diag} \left(0,\tfrac{S^{\prime \prime }(\rho)}{3S(\rho)} , -\tfrac{S^{\prime \prime }(\rho)}{6S(\rho)}, -\tfrac{1}{6} S^{\prime \prime}(\rho) S(\rho) \right)\\
 	\mathcal{B}_{ab} &:= \,^*C_{acbd} t^c t^d = 0 \; . 
 \end{align}

\section{Solutions of the  $SU(2)$ ENLSM}\label{sec:non-vanishing-baryon-current}
This section describes the dynamical spacetimes that are solutions to the full Einstein-$SU(2)$ NLSM in Eq.~\eqref{action} with a non-vanishing baryon current.
 The backreaction of the non-trivial  $\Phi=F(u)$ is 
 the piece  $H(u,\rho,\phi)$ that makes the metric a parallel wave. We present the general solution of Eq. \eqref{Heq} and single out a unique preferred one.  
  For this, we study  its asymptotic behavior, which is used throughout the rest of this section. 
  Next, we probe the spacetime geometry through the study of geodesics in Sec.~\ref{sec:geom}. 
  The baryon 
  charge is discussed in Sec.~\ref{sec:baryon-charge}. Finally,  in Sec.~\ref{sec:mass} 
  we review for the static case $F'=\w$ different notions of mass per length \textcolor{black}{and analyze its connection to the baryon charge.}

The metric is   Eq.~(\ref{background}) with $H \neq 0$, the $SU(2)$ field has $\alpha$ and $R$ as in see Eq.~(\ref{gs}), and 
\begin{equation}
\Theta= q \phi, \;\;\; \Phi=F(t-z)=F(u) \not \equiv 0.
\end{equation}
The field equation (\ref{Heq}) for $H$ is, in view of $F \not \equiv 0$, nontrivial and 
admits a solution of the form 
\begin{equation} \label{sol}
H(u,r,\phi)= -(F'(u))^2[h(r) + \psi(r) \cos(2q\phi)], 
\end{equation}
where
\begin{equation}
\begin{split}
&h''(r)= \left( \frac{\ell}{q} \right)^2\; R''(r) \exp(-2R(r))\\
&\psi''(r)-4q^2 \psi(r) = -\left( \frac{\ell}{q} \right)^2 R''(r) \exp(-2R(r)) .
\end{split}
\end{equation}

Particular solutions for these equations are:
\begin{equation}\label{h(r)2}
h(r)= \left( \frac{\ell}{q} \right)^2\int_r^{\infty} dz  \int_{z}^{\infty} e^{-2R(y)} R''(y) dy 
\end{equation}
and 
\begin{align}\nonumber
\psi(r) &= \frac{\ell^2}{4|q|^3} \left[ e^{2|q|r} \left( \int_{r}^{\infty} e^{-2|q|y -2R(y)} R''(y) dy \right)  +
e^{-2|q|r} \left( \int_{-\infty}^r e^{2|q|y -2R(y)} R''(y) dy \right) \right] \\
&=:   \psi_1(r)+\psi_2(r) . \label{spb}
\end{align}
Note that, since $R''(r)=K \kappa q^2 \sin^2(\a(r))>0$, both $h(r)$ and $\psi(r)$ (and indeed $\psi_1$ and $\psi_2$) 
are positive definite. 
To estimate the asymptotic form of $H$ for the particular solution (\ref{sol})-(\ref{spb}) we notice that
\begin{equation}\label{a1}
e^{-2R(r)} R''(r) \simeq  \a_{\pm} e^{2\b_{\pm}r} \; e^{\mp 2|q|r} \qquad \text{ as } \; \; r \to \pm \infty,
\end{equation}
where $\a_{\pm}$ are positive constants involving $C1, C_3, q, \kappa K$ and $\epsilon$.  \\

From (\ref{a1}) follows that, for Case 1 (Eq. \eqref{B}),
\begin{equation}
h(r) \simeq 
\begin{cases} 
	\frac{\a_{\scriptstyle +} \ell^2}{4 q^2(\b_+-|q|)^2} e^{2(\b_+-|q|)r} & \qquad \text{as}  \; \;    r \to \infty \\ 
	-Jr  & \qquad \text{as}  \; \;    r \to - \infty 
\end{cases} 
\end{equation}
and
\begin{equation}
\psi_1(r) \simeq \begin{cases} \frac{\alpha_+ \ell^2 }{8 |q|^3 (2|q|-\b_+)} e^{2(\b_+-|q|)r} &\qquad \text{as}  \; \;    r \to \infty \\  J_1 e^{2|q|r}  &\qquad \text{as}  \; \;    r \to - \infty \end{cases} 
\end{equation}
\begin{equation} \label{psi2a}
\psi_2(r) \simeq \begin{cases} \frac{\a_+ \ell^2}{8 |q|^3\b_+} e^{2(\b_+-|q|)r} &\qquad \text{as}  \; \;    r \to \infty \\  \frac{\a_- \ell^2}{8 |q|^3(\b_-+2|q|)}e^{2(\b_-+|q|)r}
 &\qquad \text{as}  \; \;    r \to -\infty \end{cases} 
\end{equation}
where $J$ and $J_1$ are positive constants:
\begin{equation}
J = \left( \frac{\ell}{q} \right)^2\int_{-\infty}^{\infty}  e^{-2R(y)} R''(y) dy, \qquad J_1 =  \frac{\ell^2}{4|q|^3} 
 \int_{-\infty}^{\infty} e^{-2|q|y -2R(y)} R''(y) dy.
\end{equation}
The above formulas are also valid in Case 2, with the exception of Eq.~(\ref{psi2a}):
\begin{equation}
\psi_2(r) \simeq \begin{cases}  J_2 e^{-2|q|r} &\qquad \text{as}  \; \;    r \to \infty  \\  \frac{\a_- \ell^2}{8 |q|^3(2|q|-1)}e^{2(|q|-1)r} &\qquad \text{as}  \; \;    r \to - \infty \end{cases}  \;\;\; 
\text{ (Case 2 only)}
\end{equation}
where
\begin{equation}
 J_2 =  \frac{\ell^2}{4|q|^3} 
 \int_{-\infty}^{\infty} e^{2|q|y -2R(y)} R''(y) dy.
\end{equation}

Now let us discuss the general solution of equation (\ref{Heq}). The general solution of the associated homogeneous equation is 
\begin{equation}\label{hom}
	H_h(u,r,\phi)=  A_0(u)+ A_1(u) r + \sum_{n =1}^{\infty} \left[
	\left(C_n(u) e^{nr} +D_n(u) e^{-nr}\right)\cos(n \phi) 
	+  \left(E_n(u) e^{nr} +F_n(u) e^{-nr}\right)\sin(n \phi) \right]  .
\end{equation}
Thus, the general solution of (\ref{Heq}) is $H$ given in (\ref{sol})-(\ref{spb}) plus a general solution $H_h$  above. 
The only  addition from (\ref{hom}) to (\ref{sol}) that does not worsen the general behavior as $|r| \to \infty$ 
is of the form $H_h= X F'(u)^2 r$. A suitable choice of $X$  moves the linear (in $|r|$) growth 
as $r \to -\infty$ to a linear in $r$ growth as $r \to \infty$. For this reason, in what follows we will stick 
to the particular solution in Eq.~(\ref{sol}). \\

Collecting our results we find the following behavior of $H$ in terms of $\rho$: \\

In Case 2, Eq. \eqref{C}, we  obtain
\begin{equation}
H \simeq -F'(u)^2 
\begin{cases}  
	-I \ln(\rho/\ell) &\qquad \text{as}  \; \;    \rho \to 0 \\
	C \, \cos(2q\phi)  [(\rho_\infty-\rho)/\ell]^{2|q|} &\qquad \text{as}  \; \;    \rho \to  \rho_\infty ,
\end{cases}
\end{equation}
where $C$ is a positive constant. 
This behavior is singular in both poles of the sphere. We therefore disregard this case from now on. \\

In Case 1, Eq. \eqref{B}, $H$  has the asymptotic forms
\begin{equation}\label{ABasymp}
H \simeq -F'(u)^2 
\begin{cases} 
	(-J/\beta_-) \ln(\rho/\ell) &\qquad \text{as}  \; \;    \rho \to 0 \\
	\left[ A + B \, \cos(2q\phi) \right] (\rho/\ell)^{\frac{2(\b_+-|q|)}{\b_+}} &\qquad \text{as}  \; \;    \rho \to \infty  ,
 \end{cases}
\end{equation}
where $A, B, J$ are positive constants and $\b_+-|q|$ is negative in view of (\ref{B}).
 Note that $H$ is bounded from above  (assuming, as we do, that $F'$ is bounded),  
and that it is negative definite if $B<A$.  Since $A=\frac{ \alpha_+ \ell^2}{4 q^2 (\beta_+-|q|)^2}$ and $B=\frac{ \alpha_+ \ell^2}{4 q^2 \beta_+(2|q|-\beta_+)}$, 
this is the case as long as $|q|$ is not too large. Specifically, if $1<|q|< \frac{2+\sqrt{2}+4 K\kappa}{1+8 K \kappa+8 (K \kappa)^2}$,  
$H$ is  negative everywhere (this constraint on $|q|$ uses that $K \kappa<1/(2\sqrt{2})$). \\

The behavior of the function $-H$ as a function of $\rho$ for large $\rho$ determines the causal behavior of the spacetime 
\cite{Flores:2002fx}. In our case we find from (\ref{ABasymp}) that, for large $\rho$,  
\begin{equation}
-H < -F'(u)^2 [A+B](\rho/\ell)^{\frac{2(\b_+-|q|)}{\b_+}}.
\end{equation}
Since $\frac{2(\b_+-|q|)}{\b_+}<0$ this behavior falls well in the subquadratic case   
($-H \sim \rho^p, p <2$ for large $\rho$ and fixed $(u, \phi)$) 
in the classification in \cite{Flores:2002fx}. This guarantees that the spacetime is strongly causal 
(Theorem 3.1 in \cite{Flores:2002fx}).

\subsection{Geometry of the spacetime}\label{sec:geom}

 The class of spacetimes of the form (\ref{wfw}) was studied in \cite{Candela:2002rr,Flores:2002fx,Flores:2004dr}. 
In the most interesting case where $F' \neq 0$ (and consequently $H\neq 0$), however,  our case deviates 
slightly from the one studied in the above references, because the singular behavior of $H$ as $\rho \to 0$ (see Eq.~(\ref{ABasymp})) implies that 
the spacetime manifold is not $\mathbb{R}^2_{(u,v)}\times {\mathcal{W}}$ but 
\begin{equation}\label{pw}
	(\mathbb{R}^2_{(u,v)}\times \mathcal{W}) - ( [u_1,u_2] \times \mathbb{R}_v \times  \{ p \}),
\end{equation}
where $p \in \mathcal{W}$ is  the point $\rho=0$ and $[u_1,u_2]$ is the closure of the support of $F'$. 
We will see below, however, that a large family of geodesics is indeed well defined 
in the entire  $\mathbb{R}^2_{(u,v)}\times {\mathcal{W}}$, 
as $H$ simply drops from the geodesic equation: the singularity introduced by $H$ is rather mild. \\
For a metric of the form in Eq.~(\ref{wfw}), $H$ does not contribute to any of the algebraic invariant 
scalar fields made out of the Riemann tensor, the metric, its inverse and its volume form. The metric (\ref{wfw}), however, which for $H=0$ 
is type D  in the Petrov classification, is generically type II if $H \neq 0$ 
(requiring that  it be of type D   imposes  a partial differential equation for $H$ which is 
incompatible with the field equations). As remarked above, the dominant and strong energy conditions are satisfied, and the spacetime 
is causally well behaved. \\

We proceed now to the study geodesics, for which we recall that we choose a time orientation such that 
the null vector field  $k^a \p_a =\p_v$,  which is covariantly 
constant and normal to 
the wave fronts $u=$constant, 
is 
future oriented.  
The affine geodesics are obtained from the Euler-Lagrange equations of
\begin{equation}\label{L}
	\mathcal{L} =  - \dot u  \dot v + H(u,\rho,\phi) \dot u^2  +\underbrace{\dot \rho^2 +
		S(\rho)^2 \dot \phi^2}_{=\mathcal{L}_{\mathcal{W}}} ,
\end{equation}
where a dot denotes derivative with respect to the affine parameter $s$, which is chosen such that 
\begin{equation}
	\mathcal{L} = \kappa = \begin{cases} 1 & \text{if spacelike,} \\ 0 & \text{if null,} \\ -1 & \text{if timelike.} \end{cases}
\end{equation}
Given the selected time orientation, future oriented causal curves must  satisfy
\begin{equation}\label{fut}
	\dot u \geq 0.
\end{equation}
Now let  $(x^1,x^2)=(\rho,\phi)$,  
$g_{\tiny{\mathcal{W}}}^{ij}$ 
and ${\Gamma_{\tiny{\mathcal{W}}}}^i_{jk}$, $i,j,k=1,2$ the metric inverse  and 
Christoffel symbols for  $ds^2_\mathcal{W}$. The geodesic equations from Eq.~(\ref{L}) are:
\begin{align}\label{x}
	& \ddot x^i +  {\Gamma_{\tiny{\mathcal{W}}}}^i_{jk} 
	\dot x^j \dot x^k + \Gamma^i_{uu} \dot u^2 = 0,   \\
	& \ddot v + 2 \Gamma^v_{ju} \dot x^j \dot u + \Gamma^v_{uu} {\dot u}^2=0,   \\
	& \ddot u=0, \label{ueq}
\end{align}
with 
\begin{equation}\label{GH}
\Gamma^i_{uu} = -\tfrac{1}{2} g_{\tiny{\mathcal{W}}}^{ij} \p_j H, \qquad 
\Gamma^v_{uu} = - \p_u H, \qquad \Gamma^v_{ju} = -\p_j H.
\end{equation}
From these equations follows  that  $\Gamma^*_{* v}=0$, justifying our assertion above  that $k^a$ is covariantly constant. \\
From Eq.~(\ref{ueq}), we obtain
\begin{equation}\label{solu}
u (s) = \dot u_o  s + u_o, 
\end{equation}
where $u_0$ and $\dot{u}_0$ are constants and represent the initial `position' and `velocity', respectively, of $u$ at $s=0$.
This naturally leads us to consider two different types of geodesics:

\begin{itemize}
\item $\dot u_o=0$, then $u(s)=u_o$ for all $s$. \\

For these geodesics, since $\dot u =0$, $H$ decouples from the geodesic equations (\ref{x})-(\ref{GH}), which 
then have smooth coefficients and can cross the origin at $\rho=0$ even if $u_o$ in Eq.~(\ref{dotu0}) is within the support of $F$. 
From Eqs.~(\ref{x})-(\ref{ueq}), we obtain 
\begin{equation}\label{dotu0}
(u,v,x^j) = (u_o, v=\dot v_o s + v_o, x^j(s)) ,
\end{equation}
where $x^j(s)$  is a geodesic of $\mathcal{W}$, that is, a solution of the Euler-Lagrange equations 
for the Lagrangian $\mathcal{L}_{\mathcal{W}}$ in Eq.~(\ref{L}).  \\
The only \textit{future causal geodesics} of this type  are those with constant $x^j$, that is, 
null geodesics with tangent $k^a$:
\begin{equation}\label{ofcc}
	(u_o, v=\dot v_o s + v_o, x^j(s)=x^j_o), \qquad \dot v_o>0. 
\end{equation}
This shows, in pass, that no causal closed geodesics exist in this family, since $s \to (u(s),v(s),x^j(s))$ is injective. 
The geodesics in this class with non-constant  $x^j(s)$ 
are spacelike and, if $\dot v_o=0$, they are contained in a $(u=u_o,v=v_o)$ submanifold  $\mathcal{W}$. 
These submanifolds are then totally geodesic. In particular, if $\mathcal{W}$ were incomplete (which is not our case since we have chosen $\b_-=1$),  
there would be incomplete spacetime geodesics of the form (\ref{dotu0}).

\item $\dot u_o \neq 0$, $u(s)=\dot u_o s + u_o$ (since for future causal geodesics $\dot u_o \geq 0$, and the orientation of spacelike 
geodesics is irrelevant, we will assume $\dot u_o>0$).\\

In this case, $u$ is given by Eq.~(\ref{solu}). 
Eq.~(\ref{x}) for the $x^j$ follows from a Lagrangian obtained from $ \mathcal{L}_{\mathcal{W}}$ by adding  a 
time-dependent (that is, $s$-dependent) potential:
\begin{equation}\label{el2}
\hat{\mathcal{L}}_{\mathcal{W}} = \dot \rho^2 +
S(\rho)^2 \dot \phi^2 + H(\dot u_o s +u_o,\rho,\phi) {\dot u_o}^2 .
\end{equation}
The Euler-Lagrange equations from $\widehat{\mathcal{L}}_{\mathcal{W}}$ in Eq.~(\ref{el2}), using 
$H(\dot u_os+u_o,\rho,\phi) = -{F'}^2(\dot u_os+u_o) [h(\rho)+\psi(\rho) \, \cos(2|q|\phi)]$ are 
(a prime on functions of a single variable denotes a derivative)
\begin{equation}
	\begin{split}\label{tdp}
		&2 \ddot \rho = 2 S(\rho) S'(\rho) \dot \phi^2  - \dot u_o^2 {F'}^2(\dot u_os+u_o) [h'(\rho)+\psi'(\rho) \, \cos(2|q|\phi)] , \\
		& \frac{d}{ds} (2 S(\rho)^2 \dot{\phi}) = 2|q| \dot u_o^2 {F'}^2(\dot u_os+u_o) \psi(\rho) \sin (2|q|\phi) .
	\end{split}
\end{equation}
The solutions $x^j(s)=(\rho(s),\phi(s))$  of  Eq.~(\ref{tdp}) can be 
obtained from the simpler, particular solutions  $(\tilde \rho(s),\tilde \phi(s))$ that correspond to the case 
with $\dot u_o=1$ and $u_o=0$, via
 the mapping (see Theorem 3.2 in \cite{Candela:2002rr}):
\begin{equation}\label{unitspeed}
	\begin{split}
		\rho(s)&=\tilde \rho((s-u_o)/\dot u_o),\\
		\phi(s)&=\tilde \phi((s-u_o)/\dot u_o).
	\end{split}
\end{equation}
After solving the equations for $x^j(s)$,  $v$ can be  obtained as a final step using the first integral $\mathcal{L}=\kappa$:
\begin{equation} \label{vds}
v(s) = v_o+\frac{1}{ \dot u_o} \int_{s_o}^s \left[ -\kappa +H(\dot u_o \tilde{s} +u_o,x(\tilde{s})) {\dot u_o}^2 + \mathcal{L}_{\mathcal{W}}(x(\tilde{s}),\dot x(\tilde{s})) \right] \; d\tilde{s} .
\end{equation}
In the particular case where  $F'(u)=\w \neq 0$ is a non-zero constant, the metric is stationary and consequently there is an additional constant of motion. This is 
reflected in the fact that the potential in Eq.~(\ref{el2})  is time-independent, so that the energy 
\begin{equation}
E=\dot \rho^2 +S(\rho)^2 \dot \phi^2 - H(u_o,\rho,\phi) {\dot u_o}^2
\end{equation}
is conserved. Given the behavior of $H$ as $\rho \to 0$ (see Eq.~(\ref{ABasymp})), the potential 
energy becomes infinite as $\rho \to 0$ and thus $\rho=0$ is 
\textit{unreachable}.\\

In what follows we analyze  the  more interesting case 
 of a passing wave, that is,  $F'\neq 0 $ for  $u_1 < u <u_2$ with $u_1$ and $u_2$ finite.  
In this case, the time-dependent potential is turned on only in the ``time interval'' $s_1<s<s_2$, where 
\begin{equation}\label{sj}
	s_j=(u_j-u_o)/\dot u_o, \qquad  j=1,2.
\end{equation}
In the non-trivial time interval $s_1<s<s_2$, Eq. \eqref{tdp}  admit radial solutions $\phi=\phi_o$ with $ \sin (2|q|\phi_o)=0$ and 
\begin{equation}\label{ddr}
	2 \ddot \rho =  -{F'}^2(\dot u_os+u_o) V'(\rho), \;\;\; V(\rho) \equiv h(\rho)+\psi(\rho) \, \cos(2|q|\phi_o).
\end{equation}
We would like to explore the possibility of reaching $\rho=0$ along such a radial geodesic if the geodesic was 
approaching this point when the wave arrived (i.e., $\dot \rho(s_1)<0$). It is important to keep in mind 
Eq.~(\ref{ABasymp}), which implies that $V(\rho) \simeq -J \ln(\rho/\ell)$ as $\rho \to 0$ with $J$ a positive constant.
The asymptotic behavior of $V$ as $\rho \to 0$ implies that $V'<0$ in some interval $0<\rho<\rho^*$. We assume, together with 
$\dot \rho(s_1)<0$,  
that $\rho_1=\rho(s_1) < \rho^*$.
As a result, the right hand side of Eq.~(\ref{ddr}) is non-trivial and positive for $s \in (s_1,s_2)$ so that the time-dependent potential tends to halt 
the approach to $\rho=0$. To evaluate whether this happens or not, 
 we use that $F$ has compact support, and so does $F'$. Assuming $F'$ is continuous,  it is  then necessarily  bounded. In particular, there
  is a positive $c$ such that $F'^2<c$.  This implies that the positive acceleration $\ddot \rho$ is bounded:
\begin{equation}
	0 < 2 \ddot \rho < - c V'(\rho), 
\end{equation}
then, through the interval where  $\dot \rho<0$, 
\begin{equation}
	2 \dot \rho \ddot \rho > -c V'(\rho) \dot \rho.
\end{equation}
Assuming all these conditions hold for $s_1<s<s_2' \leq s_2$ and integrating  the above inequality gives 
\begin{equation}\label{ine1}
	\dot \rho(s_2')^2 > \dot \rho(s_1)^2 + c \underbrace{\left[ V(\rho(s_1))- V(\rho(s_2'))\right]}_{<0}  .
\end{equation}
This equation guarantees that $\rho=0$ \textit{cannot} be reached for $s  \in (s_1,s_2)$, since 
$V(\rho) \to -\infty$ as $\rho \to 0$ and the available kinetic energy $\dot \rho^2$ would be entirely used up 
before this happens. 
Moreover, this analysis also allows us  to show that, \textit{for sufficiently large} $\dot u_o$, these radial geodesic can cross 
the wave without reversing the sign of $\dot \rho$, that is, $\rho(s_2)>0$ and $\dot \rho(s_2)<0$ is possible.
This will be the case if the right hand side of the inequality (\ref{ine1})  is positive for $s_2'=s_2$. Since in 
 view of Eq.~(\ref{unitspeed}),
\begin{equation}
	\begin{split}\label{eeel}
		\rho(s_1) &= \tilde \rho((s_1-u_o)/\dot u_o)= \tilde \rho((u_1-u_o)/\dot u_o^2-u_o/\dot u_o),\\
		\rho(s_2) &= \tilde \rho((s_2-u_o)/\dot u_o)= \tilde \rho((u_2-u_o)/\dot u_o^2-u_o/\dot u_o),
	\end{split}
\end{equation}
where the function $\tilde \rho$ does \textit{not} depend  on $\dot u_o$, neither on $u_o$,  then it is clear 
from (\ref{eeel}) that $\rho(s_2)$ can be made as close as we wish to $\rho(s_1)$, and the inequality  
\begin{equation}
	\dot \rho(s_1)^2 +c [V(\rho(s_1)- V(\rho(s_2))] >0
\end{equation}
is satisfied by picking $\dot u_o$ large enough. Note that, in any case, the integral defining $v(s)$ in Eq.~(\ref{vds}) is convergent. 
The  conditions $\dot \rho(s_2)<0$, $\dot \phi(s_2)=0$ guarantee that the geodesic will reach $\rho=0$ at $s=s_2- \rho(s_2)/\dot \rho(s_2)$, 
since $\dot \rho$ is a constant for $s>s_2$. \\

In summary, for passing waves,  we have found two kinds of future causal geodesics reaching (and crossing) $\rho=0$: 
the null curves of the form (\ref{ofcc}), where $u_o$ may or may not belong to the support of $F'$, and the radial causal geodesics above. 
In the latter, $u \not \in [u_1,u_2]$ 
when $\rho=0$ is crossed, the geodesic stays within the domain 
(\ref{pw}).
\end{itemize}

\subsection{Baryon charge}\label{sec:baryon-charge}

The metric induced on a $t=t_o$ hypersurface $\Sigma$,
\begin{equation} \label{Sigma}
ds^2_\Sigma = (1+H ) \; dz^2 + d\rho^2 +S^2(\rho) \; d\phi^2,
\end{equation}
is, in view of Eq.~(\ref{ABasymp}),  spacelike \textit{for sufficiently large} $\rho$. 
Given any  everywhere spacelike hypersurface $\Sigma'$ that asymptotically matches $\Sigma$, we can use 
 the results from Sec.~\ref{Spws}, specifically, Eq.~(\ref{se2}), to calculate the  baryon charge on $\Sigma'$: 
\begin{equation}
B = \int_{\Sigma'} J_a n^a \epsilon_{\Sigma'} = \frac{1}{2\pi^2} \int_\Sigma \sin^2 (\a) \sin(\Theta) \; d \a \wedge d \Theta \wedge d\Phi.
\end{equation}
What outcome should we expect for our field configuration? In the related Skyrme model on Minkowski spacetime, there are 
solutions for which the $SU(2)$ field $U$ is time independent, $U(t,\vec{x})=U(\vec{x})$, and furthermore satisfies 
$\lim_{|\vec{x}| \to \infty}U(\vec{x})=\bm{I}$, so that $U$ can be regarded as a map from a one point compactification of $\mathbb{R}^3$ 
(which is topologically $S^3$) onto $SU(2) =S^3$, and these maps carry a topological invariant winding number in $\pi_3(S^3)$. \\
In our case, however, the ansatz  in Eq.~(\ref{su2field}) \textit{forbids} the possibility that 
$U$ has a unique asymptotic limit on $t=t_o$ surfaces  (except for the trivial vacuum 
configuration  $\a=0$): even if $F$ in Eq.~(\ref{su2field}) has compact support, that is, it represents a passing wave, 
the limit of $\a$ 
at fixed $\rho$ (equivalently, fixed $r$ in Eq.~(\ref{gs})) and $|z| \to \infty$ will be a function of $\rho$, so the asymptotic values of $U$ on $\Sigma$ 
will not agree.  As a consequence, the value of $B$ --if finite-- should not be expected to be an integer; it has no topological meaning since, although $B$ is the integral 
on $\Sigma$ of the pullback of the $SU(2)=S^3$ volume form, $\Sigma$ cannot be regarded as a closed manifold. \\

While the baryon charge does not carry a topological meaning for this configuration, it remains an interesting conserved 
charge that  describes the matter content of the solution.  
In particular, contrary to what happens for the stationary solutions in \cite{Canfora:2020ppn}, the  $\W$-fronted parallel waves 
have a finite baryon number whenever the $z$-integral below is finite:
\begin{equation}\label{bn}
B =  \frac{1}{2\pi^2} \int_{-\infty}^{\infty} \sin^2(\a(r)) \; \a'(r) \; dr \,  \int_0^{2\pi} \sin(q \phi) q \; d\phi \int_{-\infty}^\infty F'(t_o-z) \, dz
= {\epsilon} \frac{\sin^2(q \pi)}{ \pi} \int_{-\infty}^\infty F'(t_o-z) dz,
\end{equation}
where we have used that, as $r$ grows   $\a:0 \to \pi$ for $\epsilon=1$, and the reverse for $\epsilon=-1$.
Note that $B=0$ for integer $q$, but if we, following the arguments in  \cite{Canfora:2020ppn},  allow $q=n+\tfrac{1}{2}, n \in \mathbb{Z}$  
then $B \neq 0$, and is finite for a step-like function with finite $\Delta F$. \\


In the stationary case $F'=\w$ (a constant), if  $q$ is an integer plus one half, we recover the infinite baryon charge in \cite{Canfora:2020ppn}, with 
\begin{equation} \label{bpl}
\frac{dB}{dz} = \epsilon \frac{\w}{\pi}, \;\;\;\; q\; \text{ half-integer}.
\end{equation}

\subsection{Mass per length in the static case}\label{sec:mass}

In the static case $F'=\w$, besides having a notion of baryon charge per length, Eq.~\eqref{bpl}, 
we can also define mass per length. This is so because the asymptotically timelike 
 vector field $t^a$ given in $(t,z,*,*)$ coordinates by $t^a \p_a =\p_t$ is Killing (since $t^a \p_a H=0$).
This implies that, for any constant $x$, the vector field (here $ T = T_{cd} g^{cd}$) 
\begin{equation}\label{Tx}
T^a = \left(T^{ab}  -\tfrac{1}{2}  \, x \,  Tg^{ab}\right) \, t_b 
\end{equation}
satisfies $\nabla_a T^a=0$. 
Once again, if $\Sigma'$ is a  timelike hypersurface that asymptotically agrees with a $t=$constant surface $\Sigma$, we can use Eq.~(\ref{se3}) to calculate
\begin{equation}\label{mx}
\int_{\Sigma'} T_a n^a \; \e^{\Sigma'}_{bcd} = \int_\Sigma \e_{abcd} T^a .
\end{equation}
The pullback onto $\Sigma$ of the 3-form dual of $T_a$ on the right hand side above   can be written, after using the first equation in (\ref{aR}), as
\begin{equation}\label{104}
\left[ K (1-x) q^2 \sin^2(\a(r)) + K \w^2 \ell^2 e^{-2R(r)}\sin^2(\a(r)) \sin^2(q\phi)\right] \;dr \wedge d\phi \wedge dz .
\end{equation}
Note that, for either $q \in \mathbb{Z}$ or $q=n+\tfrac{1}{2}, n \in \mathbb{Z}$, $\int_0^{2\pi} \sin^2(q\theta) d\theta=\pi$,  and that 
using again the first equation in Eq.~(\ref{aR}) we can calculate 
\begin{equation}\label{trick}
\int_{-\infty}^{\infty}   \sin^2(\a (r))\; dr =  \int_{-\infty}^{\infty}   \sin(\a (r))\; \frac{|\a'(r)|}{|q|} dr = 
\frac{1}{|q|} \;\int_{0}^{\pi}   \sin(\a)\; d\a =\frac{2}{|q|} .
\end{equation}
Thus, omitting the integration in $z$, the right side of Eq.~(\ref{mx}) gives an ``x-mass'' per $z-$unit: 
\begin{equation}\label{MUX}
\mu_x = 4 \pi K |q| (1-x)+ K \ell^2 \w^2 \pi \int_{-\infty}^{\infty}   e^{-2R(r)}  \sin^2(\a (r))\; dr .
\end{equation}
For $x=0$ and $\w=0$ (that is, $\Phi=F\equiv 0$, the case treated in Section \ref{Sstbc}), this calculation should reduce, in view of  Eq.~(\ref{Tx}), to that 
in Sec.~\ref{mf0}. In fact, for $x=0$ and arbitrary $\w$, the mass per z-unit (\ref{MUX}) gives 
\begin{equation}\label{mux0}
\mu_{x=0} =  4 \pi K |q| + K \ell^2 \w^2 \pi \int_{-\infty}^{\infty}   e^{-2R(r)}  \sin^2(\a (r))\; dr,
\end{equation}
which contains, in addition to the $\w=0$ 
string-like mass $4 \pi K |q| $ in Eq.~(\ref{mu}), a positive contribution proportional to $\w^2$. 
It is interesting to analyze the origin of this splitting. The  relation of the $F=0$ string-like mass with  
the winding number on the target $S^2$ was discussed in Section \ref{Sstbc}. The emergence of such a term 
in this case, where the target is $SU(2)=S^3$, can be traced back to the first term in Eq. \eqref{104}, which using Eq. \eqref{aR} 
as in (\ref{trick}), gives $\sim K (1-x) |q| \sin(\a(r)) \a'(r) dr \wedge d\phi \wedge dz$. Since integration in $z$ is omitted, we end up 
having  and integral of the pullback of an $S^2$ volume form (that of the $\Phi=$ constant 2-sphere in Eq.~(\ref{s3})). \\

Let us now analyze the $x=1$  mass per length. Twice this mass gives 
\begin{equation}
2 \mu_{x=1} = \int_{\Sigma'} \left(2T_{ab}  -  T g^{ab}\right) t^a n^b \e^{\Sigma'}_{pqr},
\end{equation}
which agrees with the Komar  mass (see, e.g., equation (11.2.10) in \cite{wald}), since 
\begin{equation}\label{Dt}
2 \mu_{x=1} = -\frac{1}{8 \pi}\int_{\p \Sigma'} \e_{a b c d} \nabla^b t^d = -\frac{1}{8 \pi}\int_{\p \Sigma} \e_{a b c d} \nabla^b t^d .
\end{equation}
From (\ref{MUX}), we find that 
\begin{equation}
2 \mu_{x=1} = 2K \ell^2 \w^2 \pi \int_{-\infty}^{\infty}   e^{-2R(r)}  \sin^2(\a (r))\; dr.
\end{equation}
When $\w=0$ this vanishes, as expected from (\ref{Dt}), since  $\nabla^b t^a=0$ in this case. 

\section{Conclusions}\label{sec:conclusion}
In this paper, we proved that the ENLSM in Eq.~(\ref{action}) that corresponds to the minimal coupling to gravity of the leading term of the low-energy effective QCD Lagrangian, admits parallel wave solutions of the form (\ref{background}),  
with non-planar wavefronts $ds^2_\W$. Asymptotically along the wavefronts (that is, as $\rho \to \infty$ in Eq.~(\ref{arv})), the $H$ function in Eq.~(\ref{background}) decays with a negative power of $\rho$ 
 and the metric approaches that of a cosmic string. As noticed in \cite{Canfora:2020ppn}, 
 non-stationary matter fields are compatible with stationary metrics: this happens if  $F(u)=\w \, (t-z)$ in Eq.~(\ref{su2field}). 
 \textcolor{black}{In this particular case}, different notions of mass per length were studied, one of them nicely splitting into 
 a cosmic-string like term and a contribution proportional to $\w^2$ (see Eq.~\eqref{mux0}). \\

There is a subcase where the matter field $U: M \to SU(2)$ has target $S^2 \subset S^3 =SU(2)$. 
\textcolor{black}{When regarded as a solution of the ENLSM (\ref{action}) with target $S^2$, this static solution is stable. It carries a topological charge 
$q \in \pi_2(S^2) \simeq \mathbb{Z}$,  and asymptotically  looks like a string with mass per 
length $\mu=4\pi K |q|$,  thereby offering an interesting example of a connection between a topological charge 
and a mass. This solution is smooth everywhere, free of conic singularities, and thus an example 
of a regular source for an asymptotically  string-like metric with a mass per length related to a topological charge.} \\

As explained in Section \ref{sec:baryon-charge}, since the $U$ field in the ansatz (\ref{su2field}) has a direction 
 dependent limit, the conserved baryon charge has no direct  topological interpretation.  
It would be  interesting to see if there are solutions of the field equations with a uniform asymptotic limit, 
non-trivial $SU(2)$ configurations that somehow generalize the connection found in Section \ref{Sstbc} 
between topological charges and notions of mass for the ENLSM with target $S^2$. 
If there is a direct link between mass and a conserved topological charge in the form of a bound,   
 gravitational radiation should naturally shut off when this bound is reached.
 Of course, besides the interest of the model 
(\ref{action}) as the minimal coupling to Einstein gravity of lowest QCD effective action, finding relationships between topological charges and 
mass notions in generic   ENLSM (\ref{nlsm}), that is, with arbitrary target manifolds, stands as an interesting problem 
by itself. \\

\acknowledgments 
We thank Fabrizio Canfora for many enlightening discussions throughout the development of  this work. 
GD is partially supported by grants No. PIP
11220080102479 (CONICET-Argentina) and No. 30720110
101569CB (Universidad Nacional de C\'ordoba).

  \end{document}